\documentclass[journal,10pt,letterpaper,final,twocolumn]{IEEEtran}%
\usepackage{amsmath}
\usepackage{amsfonts}
\usepackage{cite}
\usepackage{amssymb}
\usepackage{graphicx}
\usepackage[usenames,dvipsnames]{color}
\usepackage{epstopdf}

\providecommand{\U}[1]{\protect\rule{.1in}{.1in}}
\pdfoutput=1
\providecommand{\U}[1]{\protect\rule{.1in}{.1in}}

\newcommand{\qed}{\nobreak \ifvmode \relax \else
      \ifdim\lastskip<1.5em \hskip-\lastskip
      \hskip1.5em plus0em minus0.5em \fi \nobreak
      \vrule height0.75em width0.5em depth0.25em\fi}

\makeatother

\begin{document}

\title{Distortion Limited  Amplify-and-forward  Relay Networks and the $\epsilon$-critical Phase Transition}

\author{David E. Simmons,~\IEEEmembership{Student Member,~IEEE}, and Justin P.~Coon,~\IEEEmembership{Senior Member,~IEEE}\thanks{The authors are with the Department
of Engineering Science, University of Oxford, Parks Road, Oxford, OX1 3PJ, UK. Email:$\{$david.simmons,justin.coon$\}$@eng.ox.ac.uk}}
\maketitle
\begin{abstract}
We study amplify-and-forward (AF) relay networks operating with source and relay amplifier distortion, where the distortion dominates the noise power. The diversity order is shown to be $0$ for fixed-gain (FG) and $1$ for variable-gain (VG) if distortion occurs at the relay; if distortion occurs only at the source, the diversity order will be $1$ for both. 
With $\epsilon_\beta=N_0/\eta_\beta$ ($N_0$ the noise power, $\eta_\beta$ the distortion power at node $\beta\in\{S,R\}$, the source or relay), we demonstrate the emergence of what we call an $\epsilon$-critical signal-to-noise plus distortion ratio (SNDR) threshold (a threshold that emerges when $\min\{\epsilon_\beta\}$ becomes small) for both forwarding protocols. We show that crossing this threshold in distortion limited regions will cause a phase transition (a dramatic drop) in the network's outage probability. Thus, small reductions in the required end-to-end transmission rate can have significant reductions in the network's outage probability.
\end{abstract}


\section{Introduction}

Relaying will play a vital role in
 future vehicular-to-vehicle (V2V) communications due to the diversity gains it can offer, and its ability to extend a network's coverage area \cite{coopprotoutage,5733966,yamao2009vehicle,zhao2008extending}.
AF relaying is particularly attractive
for low latency requirements \cite{yang2009relay}, which will be prevalent in V2V communications.

OFDM systems are known to have a large peak-to-average power ratio (PAPR) making them much more susceptible to distortion due to nonlinearities.
Combining OFDM with channel fading can cause even larger PAPRs in
the received signal at the relay of OFDM networks, which may lead to more pronounced nonlinear distortion.
Although nonlinear effects are common in such networks, there is minimal literature pertaining to them. 
 In \cite{6292949}, the outage probability of
a two-hop cooperative OFDM  VG
relay system in the presence of relay nonlinearities is approximated,
while \cite{5506301} focuses on the bit-error rate of such a system when  nonlinear amplifications occur  at the source. Outage and symbol error rate  
expressions are obtained for FG  systems subject to nonlinear amplification at the relay in \cite{EuCNCoutage}. Outage probability expressions and power allocation strategies are obtained in \cite{simmons2015two} and \cite{EuCNC2}, respectively,  for two-way AF networks with nonlinearities at the relay.

\subsection*{Contributions of this work}
To the best of the authors' knowledge, there has been no work that specifically focuses on the effects of distortion limitation within the context of relaying systems.
In this paper, we   assess such effects. In particular, we show that:
\begin{enumerate}
\item For distortion limited FG networks:
\begin{enumerate}
 \item  relay distortion results in a diversity order of $0$. 
\item  \emph{no} relay distortion (but still source distortion) results in a diversity order of $1$. 
 \end{enumerate}
\item For distortion limited VG networks, the diversity order is always $1$.
\item For distortion limited FG \emph{and} VG networks, a critical outage probability SNDR threshold emerges. If the networks operate  below this threshold, outage will occur almost surely. However, crossing the threshold will cause a phase transition (a dramatic drop) in the network's outage probability. Thus, small changes in the required end-to-end transmission rate can have significant implications for the network's outage probability.
 \end{enumerate}
It is worth mentioning that point 1) contrast with known results when no distortion is present, \cite{hasna2003performance}, where it is shown that the diversity order is $1$ for both forwarding schemes.

We use $x\sim y$ to denote that $x$ is asymptotically equivalent to $y$, and $x:= y$ and $y=:x$ to denote that $x$ is defined to be equal to $y$.

\section{System Model}
\begin{figure}
	\centering{}\includegraphics[scale=0.23]{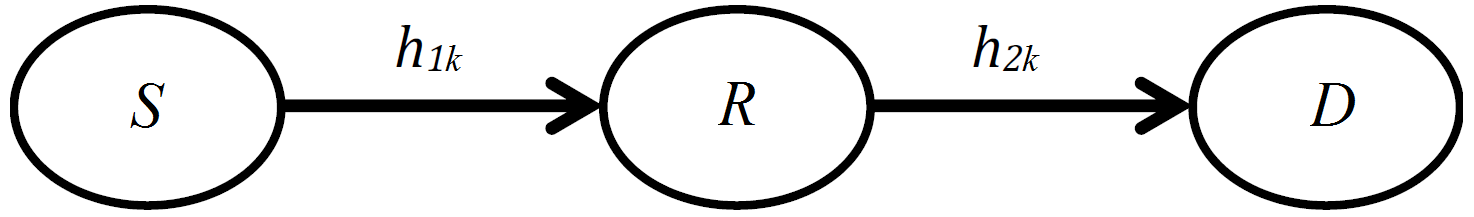}\caption{\label{fig:1}Relay network. $h_{ik}$ is the   channel coefficient for   $i$th hop on   $k$th subcarrier.}
\end{figure}

Consider a two-hop,  time-division duplexing (TDD) AF OFDM network   operating over $n$ subcarriers, Fig.~\ref{fig:1}. 
The impulse response for hop $\beta\in\{1,2\}$ is assumed to be $l$ taps long, quasi-static, and given by the time-domain (TD) vector
\begin{equation}
\bar{H}_\beta = \sqrt{\frac{n}{l}}\frac{\left[  \bar{h}_{\beta 0}\;\;  \bar{h}_{\beta 1}\;\; \cdots \;\; \bar{h}_{\beta,l-1} \right]^T}{\left|\left[  \bar{h}_{\beta 0}\;\;  \bar{h}_{\beta 1}\;\; \cdots \;\; \bar{h}_{\beta,l-1} \right]\right|},\label{eq:channelImpulseResponse}
\end{equation}
where the $i$th entry of $\bar{H}_\beta$ corresponds to the $i$th channel tap and $\tilde h_{\beta i}$, $i\in\{0,\dots,l-1\}$, are i.i.d. zero-mean complex Gaussian (ZMCG) random variables with total variance $\mu_{\beta}$. After taking the unitary Fourier transform (FT) of $\bar{H}_\beta$, the frequency response for the $k$th subcarrier of  hop $\beta\in\{1,2\}$ is given by $h_{\beta k}\sim\mathcal{CN}\left(0,\mu_\beta \right)$, $k\in\{1, \dots, n\}$. 

The relaying protocol  takes place over two  time slots. The first time-slot, amplification model, and second time-slot are detailed in the following  subsections.

\subsection{First Time Slot}

Node $S$ constructs an OFDM symbol vector  comprised of $n$ symbols, which we denote by the frequency-domain (FD) vector  $X_S=[x_{S{1}},\ldots,x_{S{n}}]^{T}$, where $\sigma_{S}^2:=\mathbb{E}[|x_{S{k}}|^{2}]$. It is assumed that the symbol  $x_{S{k}}$ is  chosen uniformly and independently from a quadrature phase-shift keying (QPSK) constellation.

\subsubsection{Cyclic Prefix Insertion and Removal}

The vector  $X_S$ is processed with an inverse FT, after which a cyclic prefix (CP) of suitable length is appended to mitigate IBI.  We assume that CP insertion and removal is performed entirely at the source and destination; i.e., not at the relay. Consequently, the CP must be more than $2l$ sample periods in length to mitigate IBI.
The TD OFDM block at  $S$ is then given by
\begin{equation}
\bar{X}_{S,cp} = \left[ \bar{x}_{S,n-2l-1}  \;\cdots \;\bar{x}_{S,n-1} \;
				  \right.  \left. \bar{x}_{S,0} \;\bar{x}_{S,1}   \cdots \;\bar{x}_{S,n-1}    \right]^T,\label{eq:SELinpSyst}
\end{equation}
where $\bar{x}_{S,i}$ is the $(i+1)$th entry of $\bar{X}_{S} := \mathbf{F}^{-1} {X}_{S} $ (the inverse FT of ${X}_{S} $). 

\subsubsection{Source Transmission}

The source has a maximum transmit power constraints, $\mathcal{P}_{max,S}$. To ensure this constraint is not exceeded, $S$ passes its TD waveform  through a soft envelope limiter (SEL) (see \cite[eq. (38)]{871400}). The output of the SEL in the TD given that the input is the $(k+1)$th element of $\bar{X}_{S} := \mathbf{F}^{-1} {X}_{S} $ is
\begin{equation}
\bar{y}_{S k}= \min\left\{\sqrt{\mathcal{P}_{max S}}, |\bar{x}_{S k}|\right\}\exp\left(i\arg \bar{x}_{S k}\right).\label{eq:aff}
\end{equation}
By considering Bussgang's theorem \cite{871400}, and provided the number of subcarriers is sufficiently large, the $k$th subcarrier frequency-domain (FD) output of the SEL can be written as
\begin{equation}
y_{S k}=\zeta_S x_{S k}+d_{S k},\label{eq:aff}
\end{equation}
where  $d_{S k}\sim\mathcal{CN}\left( 0 ,  \eta_{S}\right)$ is uncorrelated with $x_{S k}$, and $\zeta_S$ and $\eta_{S}$ are obtained from  \cite[eq. (42)]{871400}
 \begin{align}
&\zeta_\beta  =  1  -  e^{-\frac{p_{max\beta} }{  \sigma_\beta^2} }+ \sqrt{\frac{\pi p_{max\beta}}{4\sigma_\beta^2}}\mathrm{erfc}\left(\frac{ p_{max\beta} }{ \sigma_\beta^2 } \right)\label{eq:zeta_Busgang}\\
&\eta_\beta  =  \sigma_\beta^2  \left(    1  -  e^{-\frac{p_{max\beta} }{ \sigma_\beta^2} }    \right) -  \zeta_\beta^2\sigma_\beta^2.\label{eq:eta_Busgang}
\end{align}
By noting that the SEL's  input power is exponentially distributed, the average transmit power, $\mathcal{P}_{S}$, at $S$ is  obtained from 
\begin{equation}
\mathcal P_\beta =   \intop_{0}^{\infty}\frac{\min \{ x , \mathcal{P}_{max,\beta}\}e^{-\frac{x}{\sigma_{\beta}^{2}}}}{\sigma_{\beta}^{2}}dx = \sigma_\beta^2  \left(    1  -  e^{-\frac{p_{max\beta} }{ \sigma_\beta^2} }    \right).\label{eq:Tpower_Busgang}
\end{equation}
The received signal at the relay on the $k$th subcarrier is then
\begin{equation}
y_{Rk} =h_{1k}\zeta_S x_{Sk}  + v_{Rk} + h_{Ak} d_{A k},\label{eq:RecvdSigRelay}
\end{equation}
where $v_{Rk}\sim\mathcal{CN}\left(0,N_0\right)$ is a  
$k$th subcarrier noise term.

\subsection{Relay Amplification Model}

Once $y_{Rk}$, \eqref{eq:RecvdSigRelay}, has been received, the relay performs the amplification process and then transmits the resultant signal. This process takes place over two
distinct steps.
\paragraph{Step One}
The relay applies the amplification factor $G_{\alpha k}$ to the $k$th subcarrier, where $\alpha\in\{FG,VG\}$ denotes whether FG or VG is employed.
For FG, $G_{\alpha k}$ is given by
\begin{eqnarray}
G_{FGk}  &=& \sqrt{\frac{ \sigma_R^2 }{ p_A\mu_A   + N_0}} ; \label{eq:FGsystSISO}
\end{eqnarray}
where $\sigma_R^2$ is the average input power to the SEL at the relay.
 For VG, $G_{\alpha k}$ is given by
\begin{equation}
G_{VGk} =  \sqrt{\frac{ \sigma_R^2 }{p_A|h_{Ak}|^2  + N_0}}. \label{eq:VGSISO}
\end{equation}
Note, \eqref{eq:FGsystSISO} is independent of $k$. However, to aid exposition, we refrain from removing the subscript $k$. 

\paragraph{Step Two} 
The relay passes the amplified TD waveform through an SEL to limit the maximum transmit power to $\mathcal P_{maxR}$. As an immediate consequence of the central limit theorem, this TD signal converges in distribution to a stationary ZMCG variable with variance $\sigma_R^2$ as the number of subcarriers grows large. This allows us to apply Bussgang's theorem at the relay too.

It is important to note, however, that convergence to a \emph{stationary} ZMCG variable for FG is also contingent on there being a sufficient number of channel taps.
To understand this,
consider the extreme scenario in which the channels are flat across
all subcarriers; i.e., they have a single tap
response. Furthermore, note that quasi-static fading has been
considered, so the channel are  fixed
for a single OFDM TD block. For this, the subcarrier responses
will be independent of their indices, i.e., $h_{1i}=h_{1j}$ $\forall\; i,j$.
For $h_{1i}=h_1$ $\forall$ $i$, this allows us to write the TD waveform at the relay's SEL
as 
\begin{equation}
Y_{R}(t)=\frac{1}{\sqrt{n}}\sum_{k=1}^{n}e^{\frac{i2\pi kt}{n}}G_{FGk}\left(h_{1}\zeta_{S}x_{Sk}+v_{Rk}+h_{1}d_{Sk}\right).\label{eq:channeltapAVG}
\end{equation}
 Because the gains are fixed and independent of $k$, and the channel
coefficients are quasi-static, $Y_{R}(t)$ will approach
a ZMCG variable with conditional variance
\begin{equation}
\mathbb{V}\left[Y_{R}\left(t\right)|h_{1}\right]=\left(\frac{\sigma_{R}^{2}}{\mathcal{P}_{S}\mu_{1}+N_{0}}\right)\left(\mathcal{P}_{S}\left|h_{1}\right|^{2}+N_{0}\right),
\end{equation}
where conditioning is performed to account for the channel's quasi-static nature. 
Thus, we will not have the requirement for Bussgang's theorem that
the input to the SEL be a \emph{stationary} ZMCG variable with variance $\sigma_{R}^{2}$.
In particular, it will be a function of
$h_{1}$. However, as the number of channel taps grows, $h_{1i}$
and $h_{1j}$, $i\neq j$, will become increasingly decorrelated and
the averaging performed by the inverse FT in (\ref{eq:channeltapAVG})
will remove the dependence of $Y_{R}(t)$'s conditional variance
on the instantaneous realizations of $h_{1k}$, and $Y_{R}(t)$
will approach a stationary ZMCG random variable with conditional variance
$\sigma_{R}^{2}$.
Of course, in practice the number of channel taps will be finite. However,
from heuristic observations we find that $16$ or more channel
taps allows for very accurate modeling of FG systems using Bussgang's
theorem. Note,  it was shown  \cite{1045216} that a particular $20$MHz non-line-of-sight  environment would have as many as $40$ taps in its channel impulse response.

 As with before, we can  write the FD output of the SEL as
\begin{equation}
x_{Rk}=\zeta_R G_{\alpha k} y_{Rk}+d_{Rk},\label{eq:aff}
\end{equation}
where $\zeta_R$ is obtained from \eqref{eq:zeta_Busgang} and  $d_{R k}$ is uncorrelated with $y_{R k}$ and well approximated by a ZMCG random variable with variance $\eta_{R} $ obtained from  \eqref{eq:eta_Busgang}.
The average transmit power on each subcarrier at the relay is given by
$\mathcal P_{R}$ (see \eqref{eq:Tpower_Busgang}).

\subsection{Second Time Slot}
By assuming channel reciprocity, which follows from the TDD nature of the channel, and that the entire relaying process has taken place within the coherence time of the channel,  the received signal on the $k$th subcarrier at the destination is
\begin{equation}
y_{D k} = h_{2 k}x_{Rk}+v_{Dk}\label{D_B},
\end{equation}
where $v_{D k}\sim\mathcal{CN}(0,n_{0})$ is the noise term on
the $k$th carrier at  $D$. Note, to obtain \eqref{D_B} the destination  must first remove the CP from the received TD block and perform an FT on this block. The $k$th element of the FT's output vector will then be given by \eqref{D_B}.
The instantaneous per-subcarrier  SNDR  is then 
\begin{equation}
\lambda_{\alpha, k} = \frac{\sigma_S^2\zeta^2_S|h_{1k}|^2G_{\alpha ,k}^2\zeta_R^2|h_{2k}|^2}{|h_{2k}|^2\left(  G^2_{\alpha, k} \zeta^2N_{0} + \eta_S|h_{1k}|^2G_{\alpha ,k}^2\zeta_R^2   +   \eta_R\right) + N_{0}}.\label{eq:SNDR}
\end{equation}

\section{Outage Probability -   Distortion Limitation}

We begin by considering general noise and distortion conditions with the aim of quantifying the effects that distortion limitations have on performance. 

\subsection{Outage Probability Derivation}

The per-subcarrier outage probability is defined to be 
\begin{eqnarray}
P_{o , \alpha } &:=& \mathbb P\left[    \lambda_{\alpha,k}  \leq \gamma_{th}   \right]\label{eq:Po},
\end{eqnarray}
where $\gamma_{th}$ is an SNDR protection threshold.
For FG, the solution to \eqref{eq:Po} is given in \cite[eq. (16)]{EuCNCoutage} when only relay distortion occurs.  For VG, the outage probability when only relay distortion occurs can be obtained as  a special case of the two-way network given by \cite[eq. (26)]{simmons2015two}. This is done by considering the limit as the transmit power at the destination node tends to zero (i.e., when the two-way network studied in \cite{simmons2015two} becomes one-way). For VG,  \eqref{eq:Po} then becomes
\begin{eqnarray}
P_{ o,VG} \!\!\!\!\!&=&\!\!\!\!\!\left\{ \begin{array}{cc}
\!\!1 \!\!-\!\! 2 e^{-Q }  \sqrt{R }K_1\left( 2\sqrt{R }  \right)\!,&\!\! \!\!\zeta_R^2\sigma_R^2 \!>\! \gamma_{th}\eta_R\!\!\\
1,&\!\! \!\!\zeta^2_R\sigma_R^2 \!\leq\! \gamma_{th}\eta_R\!\!\\
\end{array}\right.  ,
 \label{eq:po}
\end{eqnarray}
when distortion occurs only at the relay; where $K_{1}\left(\cdot\right)$   is the first order modified Bessel function of the second kind,
\begin{eqnarray}
R:=\frac{\gamma_{th}}{\bar\Sigma_{1}\bar\Sigma_R } \left( 1     +  \frac{ \bar\Sigma_{2}\gamma_{th}}{ \bar\Sigma_{R}  }       \right); \quad
Q: = \frac{\gamma_{th}}{\bar\Sigma_{R}  }\left(  1     +   \frac{\bar\Sigma_{2}}{\bar\Sigma_{1} }    \right);\nonumber
\end{eqnarray}
and
\begin{equation}
   \bar\Sigma_{1}: =\frac{\mathcal{P}_{S}\mu_1}{N_{0}}   ,  \;
\bar\Sigma_{2} :=\frac{\mathcal{P}_{R}\mu_2}{N_{0}}, \; \bar\Sigma_{R} : =\bar\Sigma_{2} - \frac{   (1 + \gamma_{th})  \mu_2}{\epsilon_R}
\end{equation}
are the average first and second hop SNRs, and average  \emph{critical}  second-hop SNR\footnote{We use this term because  $\bar\Sigma_{R}\leq0$ implies  outage will occur almost surely.}, respectively, and $\epsilon_R:=N_0/\eta_R.$ The outage probability solution when only relay distortion occurs can be extended to the scenario where source distortion occurs  by making the substitution $\gamma_{th}\to \gamma_{th}\mathcal P_S/(\sigma^2_S\zeta^2_S - \gamma_{th}\eta_S)$, where $\sigma^2_S\zeta^2_S - \gamma_{th}\eta_S\leq0$ implies outage will occur surely. 

\subsection{Distortion Limited Performance\label{sec:distortion_limd}}

Let us consider $P_{o,\alpha}$   when distortion power dominates noise power. To do this, we make the following assumptions: 
\begin{enumerate}
\item the source power is proportional to the  relay power, 
\item the power clipping ratios,  $\mathcal P_{max,\beta} / \sigma_\beta^2$, are fixed.
\end{enumerate}
With these assumptions, $\zeta_\beta$ becomes fixed and we can write
\begin{equation}
\mathcal P_R = \tilde{\mathcal P}_R\mathcal P_S,\;\; \sigma_\beta^2 =\tilde{\sigma}_\beta^2 \mathcal P_S  \;\; \mathrm{and}\;\;\eta_\beta =\tilde{\eta}_\beta\mathcal P_S,
\end{equation}
where $ \tilde{\mathcal P}_R,\; \tilde{\sigma}_R^2,\;\tilde{\eta}_\beta\in\mathbb R^+$ are proportionality constants. 

Because of our assumptions, analyzing the system  when it is distortion limited (i.e., when $\max\{\eta_S,\eta_R\}\gg N_0$) is identical to analyzing the system in the high transmit power regime.
Consequently, we study the first order behavior of $P_{o,\alpha}$ as $\mathcal P_S\to\infty$.
With $\tilde{\sigma}_S^2\zeta^2_S - \gamma_{th}\tilde{\eta}_S>0$, for FG \eqref{eq:Po} is given to first order by \cite[Eq. (16)]{EuCNCoutage}
\begin{multline}
\!\!\!  \!\!\!P_{o,FG}   \!\sim\!   \left(  1 \! -\! e^{\!\!\!-\frac{\tilde{\eta}_R\gamma_{th}}{\left(\tilde{\sigma}_S^2\zeta^2_S - \gamma_{th}\tilde{\eta}_S\right)\tilde{\sigma}_R^2\zeta_R^2 }}  \right) \!+ \log  \left(   \mathcal P_S  \right)  \!  \left(g_{FG}  \mathcal P_S   \right)^{-1}\!;\\\label{eq:asymPoFG}
\end{multline}
while for VG, from \eqref{eq:po}, it   is given by 
\begin{eqnarray}
  P_{o,VG} \sim  \left(g_{VG} \mathcal P_S  \right)^{-1};\label{eq:asymPo}
\end{eqnarray}
where  $g_{\alpha}
  (\alpha\in\left\{ FG,VG\right\}$ 
 ) is given by
\begin{eqnarray}
\!\!\!\!g_{FG} \! \!\!&=& \!\!\!\frac{   \tilde{\mathcal P}_R\tilde{\sigma}_R^2 \zeta_R^2 \mu_B  \left( N_0\gamma_{th}\left( \tilde{\eta}_R+\tilde\sigma_R^2 \zeta_R^2 \right)  \right)^{-1}  }{    \exp\left( {-\frac{\tilde{\eta}_R\gamma_{th}}{\tilde\sigma_R^2\zeta_R^2\left(\tilde{\sigma}_S^2\zeta^2_S - \gamma_{th}\tilde{\eta}_S\right)}}   \right) }  ,  \\\;\; 
\!\!\!\!g_{VG}  \!\!\!& = &  \! \!\!  \frac{ \left( \tilde{\sigma}_R^2\zeta_R^2    \! -  \!    \frac{\tilde{\eta}_R\gamma_{th}}{\left(\tilde{\sigma}_S^2\zeta^2_S - \gamma_{th}\tilde{\eta}_S\right)}\right)\mu_1\mu_2}{N_0\gamma_{th}\left(   \mu_1    +  \tilde{\mathcal P}_R\mu_2  \right)}                   \label{eq:cgVG}
\end{eqnarray}
for FG and VG, respectively, and describes the vertical shift associated with the asymptotic outage probability curves. Assuming $\sigma^2_S\zeta^2_S - \gamma_{th}\eta_S>0$, the diversity\footnote{Strictly speaking \eqref{eq:dFG} and \eqref{eq:dVG} are not diversity gains. This is because diversity is defined in the limit as SNDR grows to infinity. With the assumptions that have been made (fixed power clipping ratios) distortion  stops the SNDR from growing without bound. However, due to their similar constuction and interpretation we refer to them as diversity gains for brevity.} of FG relaying is
\begin{eqnarray}
l_{FG}  := - \lim_{\mathcal P_S\to \infty}\left(\frac{\log  P_{o,FG}}{\log\mathcal P_S}\right)=\left\{\begin{array}{cc}0& \tilde{\eta}_R \neq 0\\1&\tilde{\eta}_R = 0\end{array} \right.\label{eq:dFG};
\end{eqnarray}
while for VG, it is independent of $\tilde{\eta}_R$, and given by
\begin{eqnarray}
l_{VG}  := - \lim_{\mathcal P_S\to \infty}\left(\frac{\log  P_{o,VG}}{\log \mathcal P_S}\right)=1.\label{eq:dVG}
\end{eqnarray}

We find that, for a fixed power clipping ratio and assuming $\sigma^2_\beta\zeta^2_\beta - \gamma_{th}\eta_\beta>0$, relay distortion only affects the diversity for FG, while distortion at the source has no affect on the diversity for either forwarding strategy. Note, the terms $g_\alpha$ \emph{are} affected by the distortion. These observations contrast with known results for FG and VG relaying \cite{hasna2003performance} where it was observed that, without nonlinear distortion at the source or relay, the diversity gains for FG and VG are always $1$.

\section{The  $\epsilon$-critical Phase Transition\label{sec:distlimit}}

To explain the results that were seen above (e.g.,  why VG asymptotically  outperforms FG with relay distortion, or why the first order expansion of the FG outage probability is lower bounded by the bracketed term of \eqref{eq:asymPoFG}), we will now define the ratio between the noise power and the distortion power at node $\beta$ to be $\epsilon_\beta$; i.e.,
\begin{equation}
\epsilon_\beta := \frac{N_0}{\eta_\beta} ,\;\beta\in\{S,R\}.\label{eq:epsilon}
\end{equation}
Furthermore, we also define $\epsilon_\star$ to be 
\begin{equation}
\epsilon_\star = \min_\beta\{\epsilon_\beta\}\label{eq:epsilonstar}
\end{equation}
With \eqref{eq:epsilon} and \eqref{eq:epsilonstar}, for FG we can write \eqref{eq:SNDR} as
 \begin{equation}
  \lambda_{FG,k}\!=\!\frac{   {\tilde{\sigma}_S^2\zeta_S^2\tilde{\sigma}_R^2\zeta_R^2|h_1|^2 }\Big/{\left(\mu_1 \max\{\eta_\beta\}\right)} }{ a_{FG} \!+\! l_{FG}\epsilon_\star  \!+\!  q_{FG} {\epsilon_\star}^2 }\label{eq:normed_outage_FG}
\end{equation}
where  $a_{FG}=   \frac{\tilde{\eta}_R}{\max\{\tilde{\eta}_\beta\}}+\frac{\tilde{\sigma}_R^2\zeta_R^2|h_1|^2\tilde{\eta}_S}{ \max\{\tilde{\eta}_\beta\}  \mu_1} $, $l_{FG}=  \frac{1}{\left|  h_{2k} \right|^2} \!+\! \frac{\tilde{\mathcal P}_R}{ \mu_1}  $, and $q_{FG} =  \frac{\tilde{\eta}_R}{ \mu_1\left|  h_{2k} \right|^2} $; and for VG  \eqref{eq:SNDR} becomes
\begin{equation}
\! \lambda_{VG,k}\! =\! \frac{    {\tilde{\sigma}_S^2\zeta_S^2\tilde{\sigma}_R^2\zeta_R^2}\Big/{ \max\{\eta_\beta\}}  }{   a_{VG} \!   +     \!    l_{VG}     \epsilon_\star   \!  +   \!   q_{VG}     \epsilon_\star  ^2 }.\label{eq:normed_outage_2}   
\end{equation}
where $a_{VG} = \frac{\tilde{\eta}_R}{\max\{\tilde{\eta}_\beta\}}+\frac{\tilde{\sigma}_R^2\zeta_R^2 \tilde{\eta}_S}{ \max\{\tilde{\eta}_\beta\}  }$, $l_{VG}=  \frac{1}{\left|  h_{2k} \right|^2} + \frac{\tilde{\mathcal P}_R}{|h_{1k}|^2}  $, and $q_{FG} =  \frac{\tilde{\eta}_R}{\left|  h_{1k} \right|^2\left|  h_{2k} \right|^2} $. 

Our goal is to analyze the system as $ \epsilon_\star \to0$. This can be done by letting  $N_0\to 0$   or $\max\{ \eta_\beta\} \to\infty$.  With the fixed power clipping ratio assumptions, both approaches will yield  
\begin{equation}
\!\!   \lambda_{FG,k}\! \sim \! \frac{\tilde{\sigma}_S^2\zeta_S^2\tilde{\sigma}_R^2\zeta_R^2|h_1|^2}{ \left( \tilde{\eta}_R\!+\! \frac{ \tilde{\sigma}_R^2\zeta_R^2 \tilde{\eta}_S |h_1|^2}{\mu_1}\right)\mu_1 } ,  
 \lambda_{VG,k} \! \sim \!     \frac{\tilde{\sigma}_S^2\zeta_S^2\tilde{\sigma}_R^2\zeta_R^2}{  \tilde{\eta}_R \!+\! {\tilde{\sigma}_R^2\zeta_R^2 \tilde{\eta}_S} }   \label{eq:eq:sim1}.
\end{equation}
From \eqref{eq:eq:sim1}, we see that the asymptotic behavior of $\lambda_{VG,k}$ is necessarily deterministic, while for $\lambda_{FG,k}$ it is deterministic only if source distortion exists and the relay is distortion free. Crucially, both of these terms are independent of $\mathcal P_S$, which is why the outage probability for FG is lower bounded (i.e., does not decay asymptotically) when relay distortion occurs, while for VG it always decays to $0$. Furthermore, the lower bound on the FG outage probability, i.e., the bracketed term of \eqref{eq:asymPoFG}, is simply the CDF of $\lambda_{FG,k}$'s asymptotic behavior.
 

\subsection{The  $\epsilon$-critical Phase Transition and SNDR thresholds}


We will now show  for FG and VG that an $\epsilon$-critical phase transition\footnote{The term `phase transition' is well established, referring to a `sudden' change in system behavior that occurs for a small change in its parameterization. The phrase `$\epsilon$
 -critical  phase transition' is novel, and the authors have chosen this as we believe it best describes itself (it is a phase transition that arrises for small $ \epsilon_\star:=\min_\beta\{\epsilon_\beta\}$)} (a phase transition arising for small $ \epsilon_\star:=\min_\beta\{\epsilon_\beta\}$) will occur about a critical value of $\gamma_{th}$, which we call the $\epsilon$-critical SNDR threshold\footnote{The phrase `$\epsilon$-critical  SNDR threshold' is novel, and we have chosen this as we believe it best describes itself. It is the critical SNDR threshold associated with the $\epsilon$-critical phase transition.}. We will then calculate the outage probability drop that occurs during this phase transition.

For FG,  consider the outage probability of $\lambda_{FG,k}$'s asymptotic behavior (see \eqref{eq:eq:sim1}), which can be manipulated into
\begin{equation}
\mathbb P\left[   \frac{ \tilde{\sigma}_R^2\zeta_R^2 |h_1|^2 }{ \mu_1\tilde{\eta}_R }  \left( \tilde{\sigma}_S^2\zeta_S^2     -     \gamma_{th}\tilde{\eta}_S  \right)  \leq \gamma_{th}       \right].\label{eq:FGDistlimLowerB}
\end{equation}
From \eqref{eq:FGDistlimLowerB}, we see that  if $\tilde{\sigma}_S^2\zeta_S^2 /\tilde{\eta}_S  \leq  \gamma_{th} $ outage will occur almost surely as $\mathcal P_S\to\infty$ (i.e., $\epsilon_\star\to0$); otherwise, it will go to the bracketed term of \eqref{eq:asymPoFG}. Consequently,  we expect  to observe a phase transition in the distortion limited region for the FG network at the critical $\gamma_{th}$ point 
\begin{equation}
\gamma_{th,c}^{(FG)}=  \tilde{\sigma}_S^2\zeta_S^2 /\tilde{\eta}_S.\label{eq:crit_thrFG}
\end{equation}
  A similar event occurs for VG, but because the asymptotic behavior of $\lambda_{VG,k}$ is necessarily deterministic, the critical  $\gamma_{th}$ point  follows immediately from \eqref{eq:eq:sim1},  and  is given by
\begin{equation}
   \gamma_{th,c}^{(VG)}   =     {\tilde{\sigma}_S^2\zeta_S^2\tilde{\sigma}_R^2\zeta_R^2}\Big/\left({  \tilde{\eta}_R + {\tilde{\sigma}_R^2\zeta_R^2 \tilde{\eta}_S} } \right)  .\label{eq:crit_thrVG}
\end{equation}
We call \eqref{eq:crit_thrFG} and \eqref{eq:crit_thrVG} the $\epsilon$-critical SNDR thresholds.

\begin{figure}
\centering{}\includegraphics[scale=0.55]{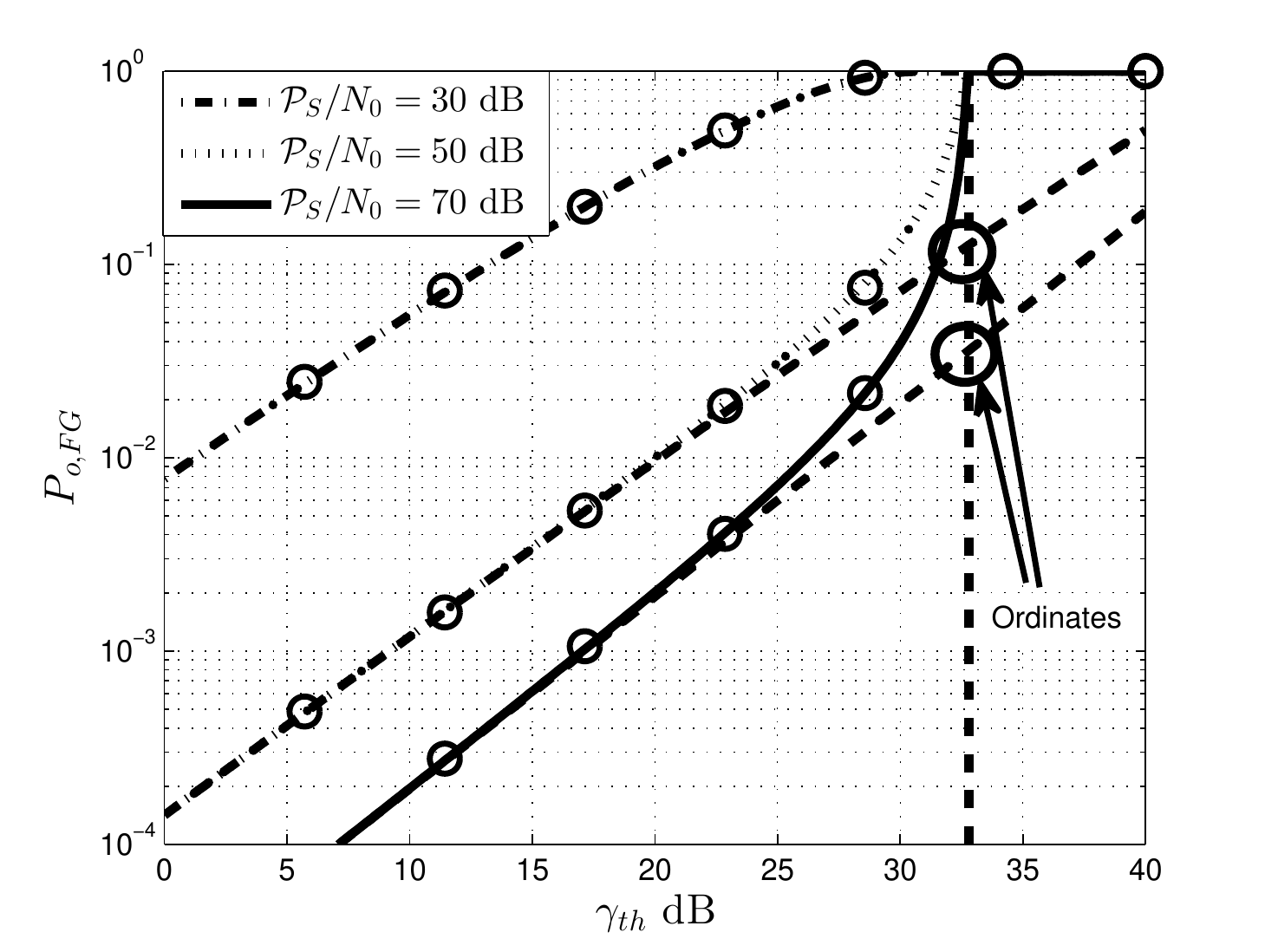}\caption{\label{fig:e1}$P_{o,VG}$ as function of $\gamma_{th}$. As $\mathcal P_S/N_0$ increases, $\epsilon_\star$ decreases and the phase transition becomes visible. We set $\mu_1 = \mu_2 = N_0 = 1$, $\mathcal{P}_{max,S}/\sigma_S^2 =5$, $\mathcal{P}_{max,R}/\sigma_R^2 =8$, $l=32$ taps. Lines - theoretical results. Markers - Monte Carlo simulations with $N=512$. Vertical dashed line is $\gamma_{th,c}^{(FG)}$, \eqref{eq:crit_thrFG}. Straight diagonal dashed lines - \eqref{eq:simSNRthFG}.   {Ordinates given (asymptotically) by \eqref{eq:OrdFG}}.}
\end{figure}

\begin{figure}
\centering{}\includegraphics[scale=0.55]{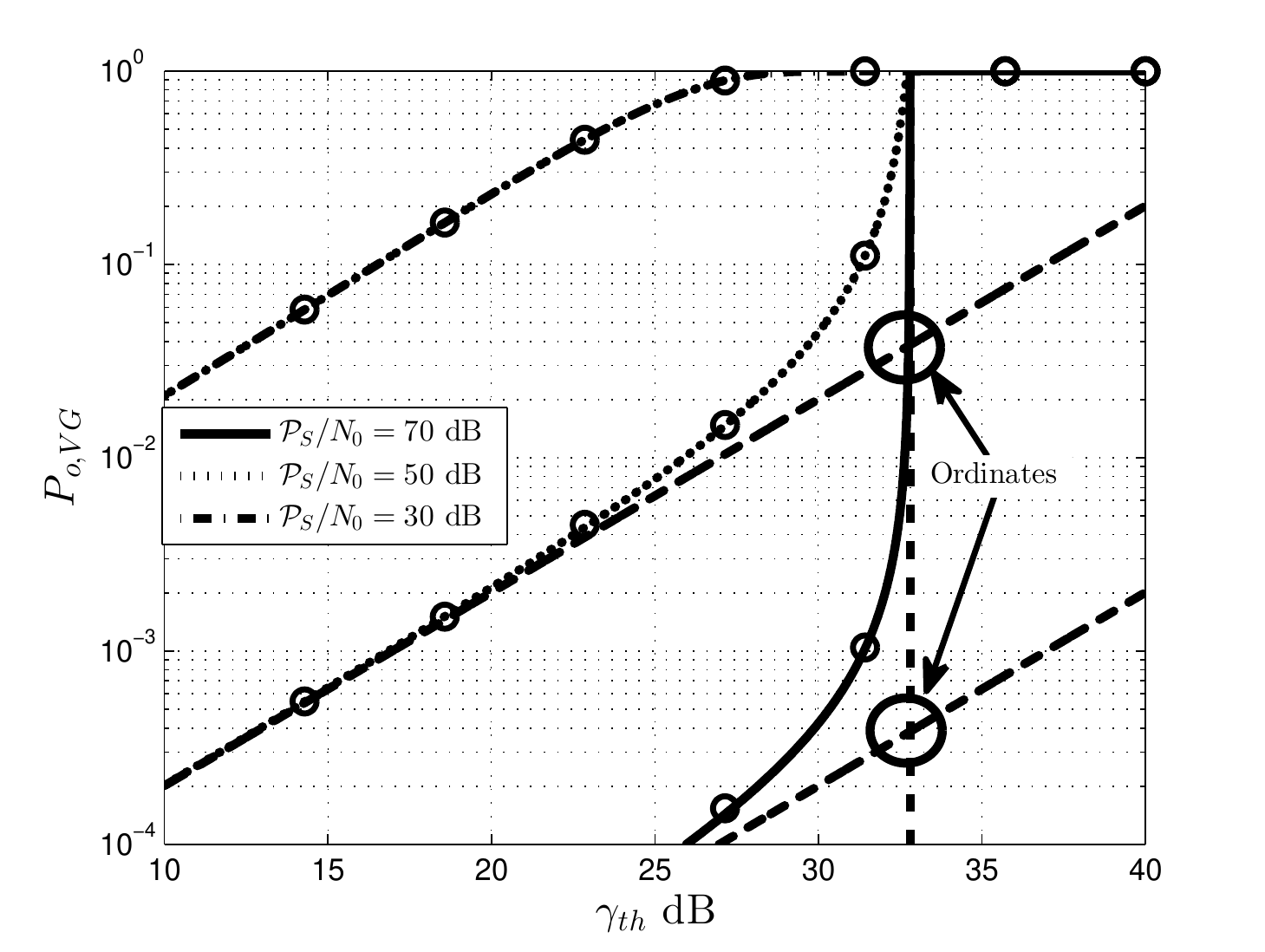}\caption{\label{fig:e2}$P_{o,VG}$ as function of $\gamma_{th}$.  As $\mathcal P_S/N_0$ increases, $\epsilon_\star$ decreases and the phase transition becomes visible.  $\mu_1 = \mu_2 = N_0 = 1$,  $\mathcal{P}_{max,S}/\sigma_S^2 =\infty$, $\mathcal{P}_{max,R}/\sigma_R^2 =5$, $l=32$  taps. Lines - theoretical results. Markers - Monte Carlo simulations with $N=512$. Vertical dashed line - $\gamma_{th,c}^{(VG)}$, \eqref{eq:crit_thrVG}. Straight diagonal dashed lines - \eqref{eq:simSNRthVG}.  {Ordinates given (asymptotically) by \eqref{eq:OrdVG}}.}
\end{figure}

Figs. \ref{fig:e1} and \ref{fig:e2} show plots of $P_{o,\beta}$ as a function of $\gamma_{th}$ for FG and VG, and the $\epsilon$-critical values of $\gamma_{th}$ (vertical dashed lines), \eqref{eq:crit_thrFG} and \eqref{eq:crit_thrVG}. From these figures, the phase transitions can be seen as $\mathcal P_S/N_0$ grows (i.e., as $\epsilon_\star$ becomes small). These figures also contain numerically generated plots (circular markers), and plots of the first order expansions (expanded about $\gamma_{th}=0$) of the outage probability functions (diagonal dashed lines). The first order expansions will be used below to establish the size of the outage probability drop occuring as $\gamma_{th,c}^{(\alpha)}$ is crossed.

\subsection{Outage Drop about the  $\epsilon$-critical SNDR Threshold}

 To establish the size of the outage probability drop about  $\gamma_{th,c}^{(\alpha)}$, we begin by calculating the first order expansions of the outage expressions at $\gamma_{th} = 0$.  For FG, this is given by
   \begin{multline}
   P_{o,FG}\! \sim\! Z  \gamma_{th}\!         \left[     \bar\Sigma_2 +  { \Sigma_1\mu_2}/{\epsilon_R }+ \right. \\  \left. \!  \left(1\!+\!\bar\Sigma_1\right)\left(1\!-\!2C\!-\!\log \left[ Z\gamma_{th} \left(1\!+\!\bar\Sigma_1\right)\right]\right)\right]     \! ,\label{eq:simSNRthFG}
\end{multline}
 where $C$ is the Euler-Mascheroni constant and $Z =  1\Big/\left({ \frac{\sigma_S^2\zeta_S^2\mu_1}{N_0} \frac{\sigma_R^2\zeta_R^2\mu_2}{N_0}}\right)$.     
For VG, it is given by 
\begin{equation}
P_{o,VG} \sim  {   Z\gamma_{th} \left(1- 2 C + \bar\Sigma_1 +   \bar\Sigma_2  -    \log\left(   Z  \gamma_{th}    \right)\right)} .\label{eq:simSNRthVG}
\end{equation}

We now use \eqref{eq:simSNRthFG} and \eqref{eq:simSNRthVG} to understand the phase transitions observed in Figs. \ref{fig:e1} and \ref{fig:e2}. This is done by evaluating them at the $\epsilon$-critical SNDR threshold (\eqref{eq:crit_thrFG} or \eqref{eq:crit_thrVG}), which will give us the ordinates illustrated in these figures and indicate the outage probability drop that should be expected as the $\epsilon$-critical threshold is crossed. For FG, with $ \tilde{\eta}_S\neq0$,  the ordinate is given to leading order about $ \epsilon_\star=0$  by
   \begin{equation}
\mathrm{Ord}_{FG} \sim       \frac{\epsilon_\star\max\{\tilde{\eta}_\beta\}}{ \tilde{\eta}_S\tilde{\sigma}_R^2\zeta_R^2\mu_2}     \left(\frac{ \mu_2}{\epsilon_R} + \log \left( \frac{\tilde{\eta}_S\tilde{\sigma}_R^2\zeta_R^2\mu_2}{     \epsilon_\star\max\{\tilde{\eta}_\beta\}     }   \right)\right)        \label{eq:OrdFG};
 \end{equation}
 while for VG, it is given by
  \begin{equation}
\mathrm{Ord}_{VG}\sim \! \frac{ \epsilon_\star^2\max\{\tilde{\eta}_\beta\}}{ a_{VG}\mu_1\mu_2}\log \left(  \frac{a_{VG}\mu_1\mu_2}{     \epsilon_\star^2\max\{\tilde{\eta}_\beta\}     }  \right).\label{eq:OrdVG}
 \end{equation}

 \subsection{A Discussion of the $\epsilon$-critical Phase Transition}
 
We will now elucidate some  observations  made in the above subsections.
First, from \eqref{eq:crit_thrFG} and \eqref{eq:crit_thrVG}, we find that
\begin{equation}
\gamma_{th,c}^{(FG)} - \gamma_{th,c}^{(VG)} =  {\tilde{\sigma}_S^2\zeta_S^2\tilde{\eta}_R}\Big/\left({\tilde{\eta}_S\tilde{\eta}_R + \tilde{\sigma}_R^2\zeta_R^2\tilde{\eta}_S}\right) .
\end{equation}
Consequently, $\gamma_{th,c}^{(VG)} \leq \gamma_{th,c}^{(FG)}$, where equality is maintained when the relay is distortion free. It follows that  if $\gamma_{th,c}^{(VG)} <\gamma_{th} < \gamma_{th,c}^{(FG)}$ FG   will outperform VG   in distortion limited regions. This is the only point at which distortion limited FG networks appear favorable over distortion limited VG networks. Moreover, if $\gamma_{th}$ belongs to this interval, by considering the FG outage probability  lower bound (the bracketed term of \eqref{eq:asymPoFG}),   FG   will outperform VG  at most by the factor
$$1-\exp\left({ - {\tilde{\eta}_R\gamma_{th,c}^{(VG)}}\Big/{\left(\tilde{\sigma}_S^2\zeta^2_S - \gamma_{th}\tilde{\eta}_S\right)\tilde{\sigma}_R^2\zeta_R^2 }}\right).$$
 
Turning our attention to the outage probability drop that occurs about $\gamma_{th,c}^{(\alpha)}$, from \eqref{eq:OrdFG} and \eqref{eq:OrdVG} we find that the ordinate of the drop scales at best (i.e., when $\epsilon_R=\infty$) like $O\left(  -\epsilon_S\log \epsilon_S\right)$ for FG, while it always scales like $O\left(  -\epsilon_\star^2 \log \epsilon_\star^2\right)$ for VG. Consequently, much larger outage probability drops should be expected for the VG. Furthermore, when $\epsilon_R\neq\infty$, \eqref{eq:OrdFG} becomes $O(1)$; i.e., it becomes independent of $\epsilon_\star$. This makes sense from our observation that the outage probability for  FG  is lower bounded when distortion occurs at the relay.

\section{Conclusion}
We studied   AF relay networks employing  nonlinearities at the source and relay while operating in distortion limited regions. 
It was shown that distortion at the source would not affect the diversity order, while distortion at the relay would affect this, but only for FG.  We then revealed a critical SNDR threshold for both FG and VG. We termed this threshold the $\epsilon$-critical SNDR threshold. Finally, it was shown that crossing the threshold in distortion limited scenarios would cause a phase transition in the   outage probability of the network.
 
 \bibliographystyle{ieeetr}
\bibliography{par}

\end{document}